\documentclass[prd,preprint,superscriptaddress,showpacs,nofootinbib,%
tightenlines ]{revtex4}
\usepackage{epsfig}
\usepackage{graphics}
\usepackage{amssymb}
\usepackage{amsmath}
\allowdisplaybreaks

\newcommand{\ben}{\begin{displaymath}}
\newcommand{\een}{\end{displaymath}}
\newcommand{\be}{\begin{equation}}
\newcommand{\ee}{\end{equation}}
\newcommand{\bea}{\begin{eqnarray}}
\newcommand{\eea}{\end{eqnarray}}
\begin{document}
\title{Vector form factor of the pion in chiral effective field theory}
\author{D.~Djukanovic}
\affiliation{Helmholtz Institute Mainz, Johannes
Gutenberg University Mainz, D-55099 Mainz, Germany}
\author{J.~Gegelia}
\affiliation
{Institut f\" ur Theoretische Physik II, Fakult\" at f\" ur Physik und Astronomie,\\ Ruhr-Universit\" at Bochum, 44780 Bochum, Germany}
\affiliation
{Tbilisi State University, 0186 Tbilisi, Georgia}
\author{A.~Keller}
\affiliation{Institute for Nuclear Physics, Johannes
Gutenberg University Mainz, D-55099 Mainz, Germany}
\author{S.~Scherer}
\affiliation{Institute for Nuclear Physics, Johannes
Gutenberg University Mainz, D-55099 Mainz, Germany}
\author{L.~Tiator}
\affiliation{Institute for Nuclear Physics, Johannes
Gutenberg University Mainz, D-55099 Mainz, Germany}
\date{October 14, 2014}
\begin{abstract}
    The vector form factor of the pion is calculated in the
framework of chiral effective field theory with vector mesons
included as dynamical degrees of freedom. To construct an
effective field theory with a consistent power counting, the
complex-mass scheme is applied.
\end{abstract}



\pacs{12.39.Fe., 11.10.Gh, 03.70.+k
}


\maketitle

\section{Introduction}

   Chiral perturbation theory (ChPT) is a well-established low-energy effective
field theory (EFT) of quantum chromodynamics in the vacuum sector \cite{Weinberg:1978kz,Gasser:1984yg}.
   The extension of this method to also include heavy degrees of freedom beyond the
Goldstone bosons is a non-trivial task, which requires both the construction of the relevant
most general Lagrangian and a suitable renormalization procedure, resulting in a self-consistent expansion
scheme for observables.
    While for the nucleon and the $\Delta(1232)$ resonance the problem of a self-consistent
momentum expansion was solved using various approaches
(see, e.g., Refs.~\cite{Bernard:2007zu,Scherer:2009bt} for a review),
the treatment of the $\rho$ meson is more complicated.
   This is mainly due to the fact that the $\rho$ meson decays into two pions, with
vanishing masses in the chiral limit.
   As a consequence of this decay mode, loop diagrams, when evaluated at energies of the order of
the $\rho$-meson mass, develop large power-counting-violating imaginary parts.
   These parts cannot be absorbed in the redefinition of the parameters of the Lagrangian,
as long as the usual renormalization procedure is used.
   Despite this feature, the heavy-particle approach was considered in
Refs.~\cite{Jenkins:1995vb,Bijnens:1996kg,Bijnens:1997ni,Bijnens:1997rv,Bijnens:1998di},
treating the vector mesons as heavy static matter fields.

   A self-consistent solution to the power-counting problem for chiral EFT with explicit
vector-meson degrees of freedom is provided by the complex-mass scheme (CMS)
\cite{Stuart:1990,Denner:1999gp,Denner:2006ic,Denner:2005fg,Actis:2006rc,Actis:2008uh,Denner:2014zga},
which is an extension of the on-mass-shell renormalization scheme to unstable particles.
   As applications of this approach in chiral EFT with heavy degrees of freedom,
the masses and widths of the $\rho$ meson and the Roper resonance were discussed
in Refs.~\cite{Djukanovic:2009zn,Djukanovic:2010zz}, respectively, as well as
electromagnetic properties in Refs.\  \cite{Bauer:2012at,Djukanovic:2013mka,Bauer:2014cqa}.
    Different approaches to the inclusion of spin-1 fields have been discussed in,
for example, Refs.~\cite{Rosell:2004mn,Bruns:2004tj,Bruns:2008,Leupold:2009nv,Terschluesen:2010ik}.

   In the present work, we consider the vector form factor of the pion in the
time-like region up to $q^2\sim 1\, {\rm GeV^2}$ in chiral EFT with vector
mesons as dynamical degrees of freedom using the CMS.
   Historically, the existence of a neutral vector meson with isospin
zero---nowadays called the $\omega$ meson---was predicted by Nambu \cite{Nambu:1957vw}
to explain the electromagnetic structure of the nucleon.
   An isoscalar piece was needed to compensate the contribution to the
mean square charge radii originating from the pion cloud.
   Shortly afterwards, Frazer and Fulco \cite{Frazer:1959gy} realized
that, within a dispersion-theoretical treatment of the form factors,
an isovector resonance would explain some features of the isovector
electromagnetic form factors of the nucleon.
   The concept of the $\rho$-meson dominance model of the pion form factor
was established by Gell-Mann and Zachariasen
\cite{GellMann:1961tg}.
   For an overview of the vector-meson dominance hypothesis, see
Refs.~\cite{Sakurai:1969,Feynman:1972}.
   In recent years, the pion vector form factor has attracted considerable
interest, in particular because of its impact on the determination of the hadronic contribution
to the anomalous magnetic moment of the muon \cite{Actis:2010gg}.
   From the theoretical side, numerous descriptions of the pion vector form factor
exist.
   For example, in Ref.\ \cite{Brandt:2013dua} the pion vector form factor has been studied
in the space-like region within lattice QCD and
next-to-next-to-leading-order ChPT, while a
new approach to the parametrization of the pion vector form factor
has been presented in Ref.~\cite{Hanhart:2012wi}.

   In this work, we fit the parameters of the effective theory to the $\tau$ decay and
describe the pion form factor data.
   However, to describe the data from $e^+ e^-\to \pi^+ \pi^-$
process we need to take into account the isospin symmetry breaking.
   This is done by including the $\rho^0$-$\omega$-$\gamma$ mixing.

\section{Lagrangian}
   To begin with, we specify the Lagrangian of pions ($\pi_a$) and $\rho$ mesons
($\rho^\mu_a$) relevant for the calculation of the vector form factor of the pion
\cite{Djukanovic:2009zn,Ecker:1989yg}:
\begin{align}
{\cal L} & =\frac{F^2}{4}\,{\rm Tr} \left[D_\mu U\left(D^\mu U\right)^\dagger\right]
+\frac{F^2\,M^2}{4}\,{\rm Tr} \left(U^\dagger+U\right)\nonumber\\
&\quad +i\,\frac{l_6}{2}\,\text{Tr}\left[f_{R\mu\nu}D^\mu U(D^\nu U)^\dagger
+f_{L\mu\nu}(D^\mu U)^\dagger D^\nu U\right]
\nonumber\\
&\quad-\frac{1}{2}\,{\rm Tr}\left(\rho_{\mu\nu}\rho^{\mu\nu}\right)
+\left[ M_{\rho}^2+\frac{c_{x}\,M^2\,{\rm Tr} \left(U^\dagger+U\right) }{4}\right]
{\rm Tr}\left[\left(\rho_\mu-\frac{i\,\Gamma_\mu}{g}\right)
\left(\rho^{\mu}-\frac{i\,\Gamma^\mu}{g} \right)\right]\nonumber\\
&\quad
+i\,d_x {\rm Tr}\left[\rho^{\mu\nu}\Gamma_{\mu\nu}\right]
- \frac{f_V}{\sqrt{2}}{\rm Tr}\left\{\rho_{\mu\nu}f^{\mu\nu}_+\right\}+\cdots,
\label{LagrangianVCh}
\end{align}
where the individual elements are defined as
\begin{align}
U&=u^2={\rm exp}\left(\frac{i\tau_a\pi_a}{F}\right),\nonumber\\
D_\mu U&=\partial_\mu U -i v_\mu U+i U v_\mu,\nonumber\\
f_{R\mu\nu}&=f_{L\mu\nu}=\partial_\mu v_\nu-\partial_\nu v_\mu,\nonumber\\
\rho_\mu&=\frac{\tau_a \rho_{a\mu}}{2},\nonumber\\
\rho_{\mu\nu}&=\partial_\mu\rho_\nu-\partial_\nu\rho_\mu - i g\left[\rho_\mu,\rho_\nu\right],\nonumber\\
\Gamma_\mu&=\frac{1}{2}\left[u^\dagger\partial_\mu u+u\partial_\mu u^\dagger
- i\left(u^\dagger v_\mu u +u v_\mu u^\dagger\right)\right],\nonumber\\
\Gamma_{\mu\nu}&=\partial_\mu \Gamma_\nu-\partial_\nu \Gamma_\mu+[\Gamma_\mu,\Gamma_\nu],\nonumber\\
f_{+\mu\nu}& =u (\partial_\mu v_\nu - \partial_\nu v_\mu) u^\dagger
+ u^\dagger (\partial_\mu v_\nu - \partial_\nu v_\mu) u.
\label{somedefinitions}
\end{align}
   In Eq.~(\ref{LagrangianVCh}), the ellipses stand for terms containing more fields
and higher orders of derivatives.
    In fact, at the beginning all the fields and parameters of Eqs.~(\ref{LagrangianVCh})
and (\ref{somedefinitions}) should be regarded as bare quantities which are usually indicated
by a subscript 0.
   However, to increase the readability of the expressions we have omitted this index.
   The external electromagnetic four-vector potential ${\cal A}_\mu$
enters into $v_\mu= - e\,{\cal A}_\mu \tau_3/2$
[$e^2/(4\pi)\approx 1/137, e>0$].
   In Eq.~(\ref{LagrangianVCh}), $F$ denotes the pion-decay constant in the chiral
limit, $M^2$ is the lowest-order expression for the squared pion mass,
$M_\rho$ is the $\rho$-meson mass in the chiral limit, $g$, $c_x$, $d_x$, and $f_V$
are coupling constants.
   Demanding that the dimensionless and dimensionfull couplings are
independent, the consistency condition for the $\rho\pi\pi$
coupling \cite{Djukanovic:2004mm} leads to the Kawarabayashi-Suzuki-Riazuddin-Fayyazuddin
(KSRF) relation \cite{Kawarabayashi:1966kd,Riazuddin:sw},
\begin{equation}
M_\rho^2= 2\,g^2 F^2.
\label{M0}
\end{equation}

    To carry out the renormalization, we use the CMS, which we implement by the following
substitution in the effective Lagrangian:
\begin{align}
\rho^\mu_0 & = \sqrt{Z_\rho}\,\rho^\mu\,, \ \ \ Z_\rho = 1 + \delta Z_\rho\,,\nonumber\\
\pi^a_0 & = \sqrt{Z_\pi}\,\pi^a\,, \ \ \ Z_\pi = 1 + \delta Z_\pi\,,\nonumber\\
M_{\rho0} & = M_R +\delta M_R \,,
\nonumber\\
c_{x 0}\left(1 + \delta Z_\rho\right) & = c_x+\delta c_x\,,\nonumber\\
g_0 & = g+\delta g\,,\nonumber\\
F_0 & = F+\delta F\,,\nonumber\\
d_{x0} & = d_x+\delta d_x\,,\nonumber\\
f_{V0} & = f_V+\delta f_V .
\label{renormparameters}
\end{align}

    We choose the renormalized mass of the vector meson as the pole of the
propagator in the chiral limit, $M_R^2=(M_\rho- i \Gamma/2)^2$.
   The loop expansions of $\delta Z_\rho$, $\delta Z_\pi$, $\delta M_R$, $\delta c_x$, $\delta g$,
$\delta F$, $\delta d_x$, and $\delta f_V$
generate counter terms.
   We include $M_R^2$ in the $\rho$-meson propagator and treat the counter terms perturbatively.
   The finite parts of the counter terms are fixed such that the loop diagrams with external
vector mesons are subtracted {at their complex ``on-shell'' points in the chiral limit, specified
by the pole position of the vector meson propagator}.

   The power-counting rules turn out to be more involved than in standard ChPT in the vacuum sector.
We use the rules specified in Ref.~\cite{Djukanovic:2009zn}.
   To determine the chiral order of a given diagram, we need to consider all
possible flows of the external momenta through the internal lines of the diagram.
   Counting the powers assigned to the propagators and vertices discussed below, we then
determine the chiral order for each flow of external momenta.
   The chiral order of the diagram is defined as the smallest amongst these orders.

   Let $q$ generically denote small quantities with the dimension of a mass such
as the pion mass, which we count as ${\cal O}(q^{1})$.
   The property small is with reference to a scale, which we take to be the mass
of the $\rho$ meson ($\sim 770$ MeV), and which we count as ${\cal O}(q^{0})$.
   The width of the $\rho$ meson counts as ${\cal O}(q^{1})$.
   Pion propagators that do not carry large external momenta count as ${\cal O}(q^{-2})$,
whereas pion propagators carrying large momenta count as ${\cal O}(q^{0})$.
   In contrast, a vector meson propagator not carrying a large external momentum counts as
${\cal O}(q^{0})$, and as ${\cal O}(q^{-1})$ if it carries a large external momentum.
   Vertices generated by the effective Lagrangian of Goldstone bosons ${\cal L}_\pi^{(n)}$
count as ${\cal O}(q^n)$ if no large external momenta are flowing through them and as
${\cal O}(q^0)$ otherwise.
   Finally, a loop integration in $n$ dimensions counts as ${\cal O}(q^n)$.

\begin{figure}
\epsfig{file=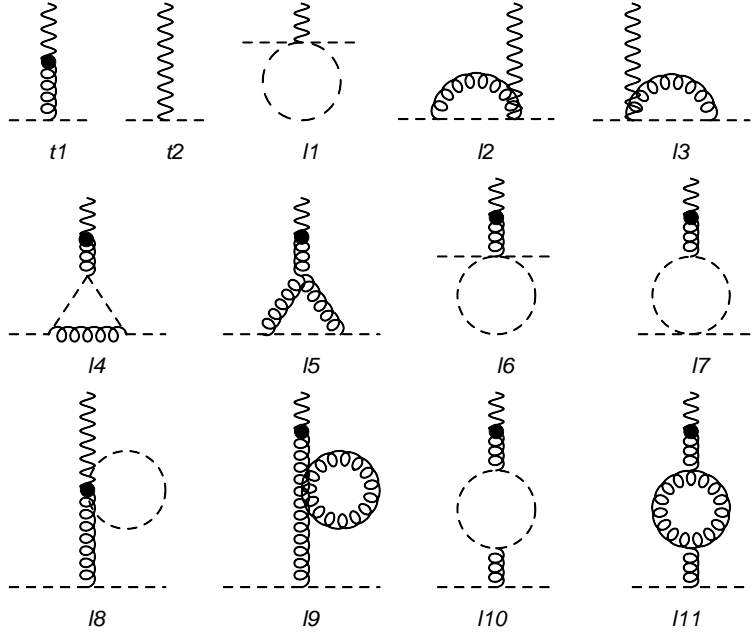, width=0.6\textwidth} \caption[]{\label{TLdiagrams:fig}
Tree and one-loop diagrams contributing to the electromagnetic form
factor of the pion. The dashed, curly, and wiggly lines
correspond to pions, vector mesons, and the vector source,
respectively.}
\end{figure}


\section{Pion form factor}

    At one-loop order, the pion form factor is given by the following expression:
\begin{equation}
F(q^2)=F_{\rm tree}\,(1+\delta Z_\pi)+F_{\rm 1 loop}\,,
\label{piff1loopgeneral}
\end{equation}
where $F_{\rm tree}$ and  $F_{\rm 1 loop}$ are the contributions of the tree and one-loop diagrams,
respectively, and $1+\delta Z_\pi$ is the wave function renormalization constant of the pion at
one-loop order.
   The explicit expression for $\delta Z_\pi$ is given in the appendix.

   The tree-level contributions to the electromagnetic form factor of the pion are shown in diagrams
$t1$ and $t2$ of Fig.~\ref{TLdiagrams:fig}.
   The corresponding expressions are given by
\begin{eqnarray}
\label{treeff1}
D_{t1} & = & \frac{\left(c_x M_\pi^2+M_R^2
-g d_x q^2\right) \left(c_x M_\pi^2+M_R^2-g D_x q^2\right)}{2
F_\pi^2 g^2 \left(c_x M_\pi^2+M_R^2-q^2\right)}\,,\\
\label{treeff2}
D_{t2} & = &\frac{2 F_\pi^2 g^2+2 q^2 l_6 g^2-c_x M_\pi^2-M_R^2}{2 F_\pi^2 g^2},
\end{eqnarray}
where $D_x = d_x-\sqrt{2} f_V.$

    The one-loop contributions to the pion form factor relevant for this work
are shown in diagrams $l1-l11$ in Fig.~\ref{TLdiagrams:fig}. 
 The corresponding
expressions are given in the appendix.

\section{$\rho^0$-$\omega$-$\gamma$ mixing}

   As emphasized in Ref.~\cite{Jegerlehner:2011ti}, the $\rho^0$-$\omega$-$\gamma$ mixing plays an important role in
describing the pion form factor extracted from $e^+ e^-\to \pi^+\pi^-$ data.
   Within the formalism of QFT, the above mixing is taken into account by solving a system of coupled
equations for the dressed propagators.
   We parameterize the proper self-energy contributions as
\begin{equation}
i\,\Pi_{xy}^{\mu\nu}(p)=i \left[\Pi_{1,xy}(p^2) g^{\mu \nu}+\Pi_{2,xy}(p^2) \,p^\mu p^\nu\right],
\label{SExy}
\end{equation}
where $x$ and $y$ stand for either $\rho$, $\omega$ or $\gamma$, and solve the system of equations for the
dressed propagators.
   The dressed propagator is given by the solution to the equation
\begin{eqnarray}
S^{\alpha\beta}_{xy}(p)=S_{0,xy}^{\alpha\beta}(p) - S^{\alpha
\gamma}_{0,xv}(p) \Pi^{\gamma\delta}_{vw}(p) S^{\delta
\beta}_{wy}(p),
\end{eqnarray}
where the matrix containing the undressed propagators is given by
\begin{equation}
S_0^{\alpha\beta}(p)=
\left(
\begin{matrix}
S_{0,\rho}^{\alpha\beta}(p) & 0 & 0 \\
0 & S_{0,\omega}^{\alpha\beta}(p) & 0 \\
0 & 0 & S_{0,\gamma}^{\alpha\beta}(p)
\end{matrix}
\right),
\end{equation}
and
\begin{align}
\label{eq:bare_propagators}
S^{\alpha\beta}_{0,\rho/\omega}(p)&=- \frac{1}{p^2-z_{\rho/\omega}^2}\left(g^{\alpha\beta}-\frac{p^\alpha p^\beta}{z_{\rho/\omega}^2}\right),\\
S^{\alpha\beta}_{0,\gamma}(p)&=-\frac{1}{p^2}\left(g^{\alpha\beta}-\frac{p^\alpha p^\beta}{p^2}\right),
\end{align}
where $z_{\rho/\omega}^2$ denotes the position of the pole of the
dressed $\rho$- or $\omega$-meson propagator.\footnote{
In the complex-mass scheme, the undressed propagator involves,
strictly speaking, the position of the pole of
the dressed propagator in the chiral limit. For the present
calculation, the difference between using the physical position of
the pole instead of its chiral limit results in higher-order terms.}
   Neglecting the $\gamma$-$\omega$ mixing, the dressed propagator of the $\rho^0$ meson has the form
\begin{equation}
S^{\alpha\beta}_{\rho^0}(p):=S^{\alpha\beta}_{\rho\rho}(p)=-\left[ g^{\alpha \beta}
\,D^1_{\rho\rho}(p^2)+ p^{\alpha
} p^{\beta }\,D^2_{\rho\rho}(p^2)\right],
\end{equation}
where
\begin{equation}
\label{D1rhorho}
D^1_{\rho\rho}(p^2)=\frac{N(p^2)}{D(p^2)}\,,
\end{equation}
with
\begin{align}
\label{Np2}
N(p^2) & =- \left[p^2-\Pi_{1,\gamma\gamma}(p^2)\right]
\left[p^2-z_\omega^2-\Pi_{1,\omega\omega}(p^2)\right],\\
\label{Dp2}
D(p^2)& =-(p^2)^3
+\left[z_{\rho}^2+z_{\omega }^2+\Pi_{1,\gamma\gamma}\left(p^2\right)
+\Pi_{1,\rho \rho }\left(p^2\right)+\Pi_{1,\omega \omega }\left(p^2\right)\right](p^2)^2 \nonumber \\
&\quad-\left\{\left[\Pi_{1,\gamma\gamma}\left(p^2\right)+\Pi_{1,\omega \omega}\left(p^2\right)\right]z_{\rho }^2
-\Pi_{1,\rho \gamma}^2\left(p^2\right)-\Pi_{1,\rho \omega}^2\left(p^2\right)\right.\nonumber \\
&\quad+\left. z_{\omega }^2 \left[z_{\rho }^2+\Pi_{1,\gamma\gamma}\left(p^2\right)
+\Pi_{1,\rho \rho}\left(p^2\right)\right]
+\Pi_{1,\rho \rho}(p^2)\Pi_{1,\omega \omega}(p^2)\right.\nonumber\\
&\quad+\left.\Pi_{1,\gamma\gamma}(p^2)\left[\Pi_{1,\rho\rho}(p^2)+\Pi_{1,\omega\omega}(p^2)\right]\right\}p^2
-z_{\omega}^2\Pi_{1,\rho\gamma}^2(p^2)\nonumber\\
&\quad-\Pi_{1,\gamma\gamma}(p^2)\Pi_{1,\rho\omega}^2(p^2)
+z_{\rho }^2 z_{\omega }^2 \Pi_{1,\gamma\gamma}(p^2)
+z_{\omega }^2 \Pi_{1,\gamma\gamma}(p^2)\Pi_{1,\rho\rho}(p^2)\nonumber\\
&\quad+\left[\Pi_{1,\gamma\gamma}(p^2)\Pi_{1,\rho \rho }(p^2)
-\Pi_{1,\rho\gamma}^2(p^2){}\right]\Pi_{1,\omega\omega}(p^2)
+z_{\rho}^2\Pi_{1,\gamma\gamma}(p^2)\Pi_{1,\omega\omega}(p^2).
\end{align}
   We do not give the explicit form of $D^2_{\rho\rho}(p^2)$, because due to the current conservation it does not contribute
to the calculation of the form factor of the pion.

   In the following, the $\rho^0$-$\omega$-$\gamma$ mixing is only taken into account
at tree level.
   This amounts to putting $\Pi_{1,xx}(p^2)$ to zero in Eqs.~(\ref{Np2}) and
(\ref{Dp2}), and keeping only $\Pi_{1,\rho\gamma}$ and $\Pi_{1,\rho\omega}$:
\begin{displaymath}
D^{1,\text{tree}}_{\rho\rho}(p^2)=\frac{p^2(p^2-z_\omega^2)}{(p^2)^3-(p^2)^2(z_\rho^2+z_\omega^2)+p^2[-\Pi_{1,\rho\gamma}^2(p^2)
-\Pi_{1,\rho\omega}^2(p^2)+z_\omega^2 z_\rho^2]+z_\omega^2\Pi_{1,\rho\gamma}^2(p^2)}.
\end{displaymath}
   For a transverse self energy, we define the functions $\Pi_{\rho \gamma}$ and $\Pi_{\rho \omega}$ by
$\Pi_{1,xy}(p^2)=- p^2 \Pi_{2,xy}(p^2)=p^2\Pi_{xy}(p^2)$.
   In fact, at tree level the functions $\Pi_{\rho \gamma}$ and $\Pi_{\rho \omega}$ are constants
and we denote them as mixing parameters.
   We allow the renormalized mixing parameters to become complex, thus incorporating the contributions of
the loop diagrams in the renormalization of the mixing parameters.
   Finally, by substituting $-D^{1,\text{tree}}_{\rho\rho}(p^2)$ for $1/(c_x M_\pi^2+M_R^2-q^2)$,
in Eq.~(\ref{treeff1}), we obtain the following expression for the tree-level diagrams:
$$
1+\frac{1}{2 F_\pi^2 g^2} \left[\frac{\left(z_\omega ^2-q^2\right) \left(z_\rho^2-g d_x  q^2\right)
   \left(z_\rho^2-g D_x q^2\right)}{(q^2)^2 \left(1-\Pi_{\rho\gamma}^2-\Pi_{\rho \omega}^2\right)
-q^2 \left[z_\rho^2+z_\omega ^2 \left(1-\Pi_{\rho\gamma}^2\right)\right]
   +z_\rho^2 z_\omega ^2}+2 g^2 l_6 q^2-z_\rho^2\right].
$$

\renewcommand{\arraystretch}{1.4}
\begin{table}
\label{fittable}
\begin{center}
\begin{tabular}{|c|c|c|c|c|c|c|c|c|}
   \hline
   {\rm Fit} &  $d_x \times 10^{-2}$ & $l_6 \times 10^{-4}$ & $\Pi_{\rho \omega} \times 10^{-2}$ & $\Pi_{\rho \gamma } \times 10^{-2}$  &  $M_\rho $ [{\rm GeV}] & $\Gamma$ [{\rm GeV}] & $M_\omega $ [{\rm GeV}] & $\chi^2_{\rm red} ({\rm dof})$   \\
    \hline
  1  & $-2.98(4)$ & 3.4(4) & $0.90(3)-i 1.38(3)$ & $-6.7(5)-i 1.6(4)$ & $0.7621(3)$ & $0.1421(5)$ & $- $        & $1.54(88)$  \\
  2  & $-2.95(4)$ & 3.4(4) & $1.05(4)-i 1.34(3)$ & $-6.6(6)-i 1.4(4)$ & $0.7622(3)$ & $0.1419(5)$ & $0.7838(2)$  & $1.32(88)$  \\
   \hline
 \end{tabular}
\end{center}\caption{
Fit parameters for the simultaneous fits of the pion
vector form factors. Data is taken from
\cite{Fujikawa:2008ma,Ambrosino:2010bv}.}
\label{fits}
\end{table}

\section{Fits}\label{fit_section}

   We perform simultaneous fits of the coupling constants and the complex mixing parameters
$\Pi_{\rho\gamma}$ and  $\Pi_{\rho\omega}$ to the $\tau$ decay \cite{Fujikawa:2008ma} and
$e^+e^-$ scattering data \cite{Ambrosino:2010bv}, where we use a range in
$q^2$ up to 1.125 $\mathrm{GeV}^2$ and $0.845$ $\mathrm{GeV}^2$, respectively.
   For the pion mass and the pion decay constant we use $M_\pi=0.1395$ GeV and $F_\pi=0.0922$ GeV.
   Moreover, we make use of the KSRF relation [Eq.~(\ref{M0})] to eliminate $g$, and set
$D_x=0$ as suggested in Ref.~\cite{Djukanovic:2005ag}.
The loop diagrams are subtracted at the physical pion mass, instead of being subtracted at chiral limit.
This eliminates numerical instabilities and the difference is of higher order for the calculation at hand.
   The coupling $g$ always appears quadratically except for the combination $g d_x$ and
$g D_x$ in the tree-level contribution.
   Our result for $d_x$ corresponds to a positive value of $g$.

   In the first fit, we fix $z_\omega=(0.7827-i\,0.0085/2)$ GeV.
   In the second fit, we allow for a floating $\omega$ mass, resulting in
an improved description with only a modest change  of 1.1 MeV in the $\omega$ mass.
   In Tab.~\ref{fits}, we show the results for the fit parameters obtained for these two fits.
   The fitted values for the $\rho$ mass and the width are consistent with earlier determinations
of the $\rho$-meson pole parameters, e.g. Ref.~\cite{SanzCillero:2002bs}.
   The results for the pion form factor are plotted in Fig.~\ref{fits} together with the
experimental data and the form factor at tree order for the same values of the parameters.



\begin{figure}
  \epsfig{file=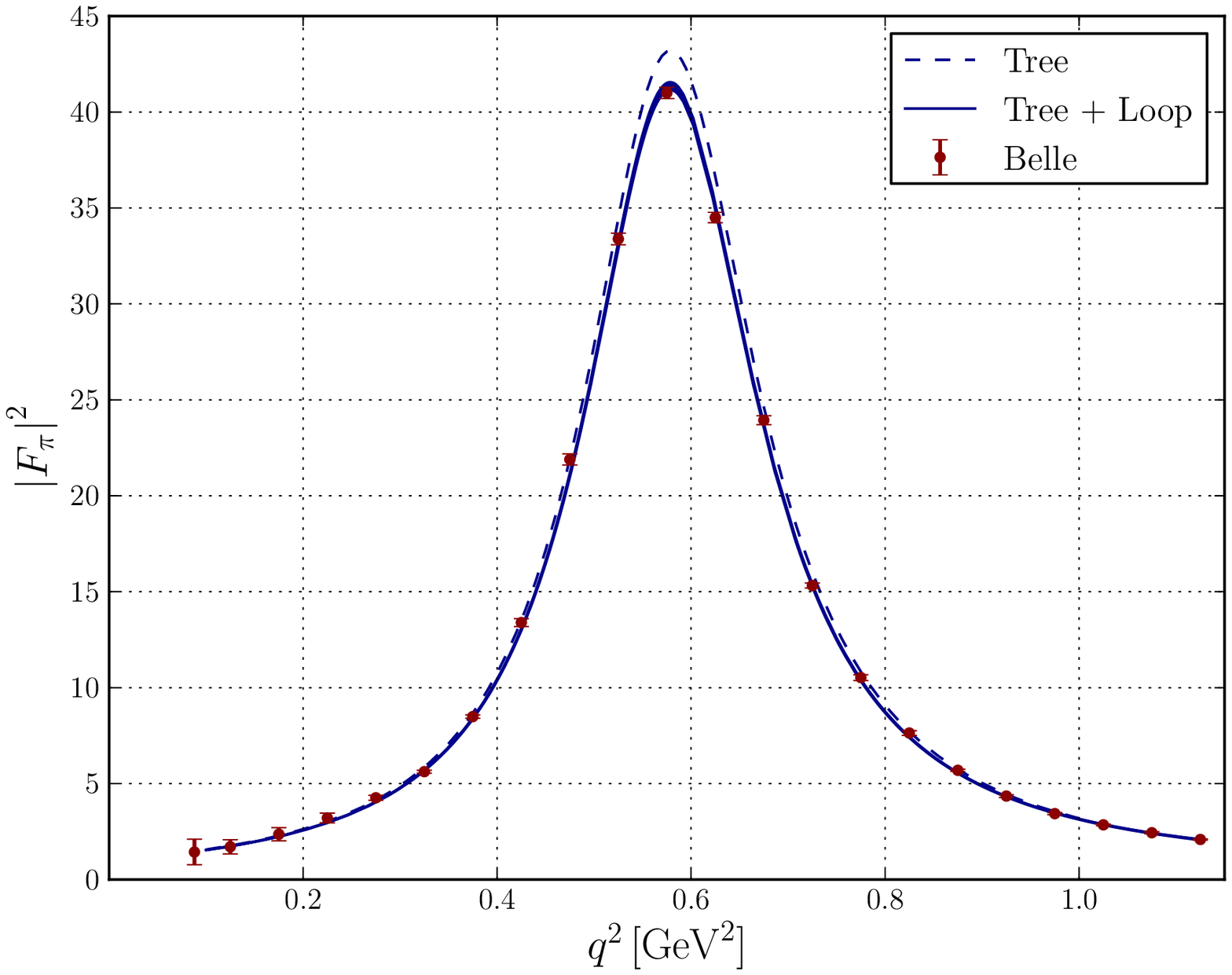,width=.4\textwidth}\epsfig{file=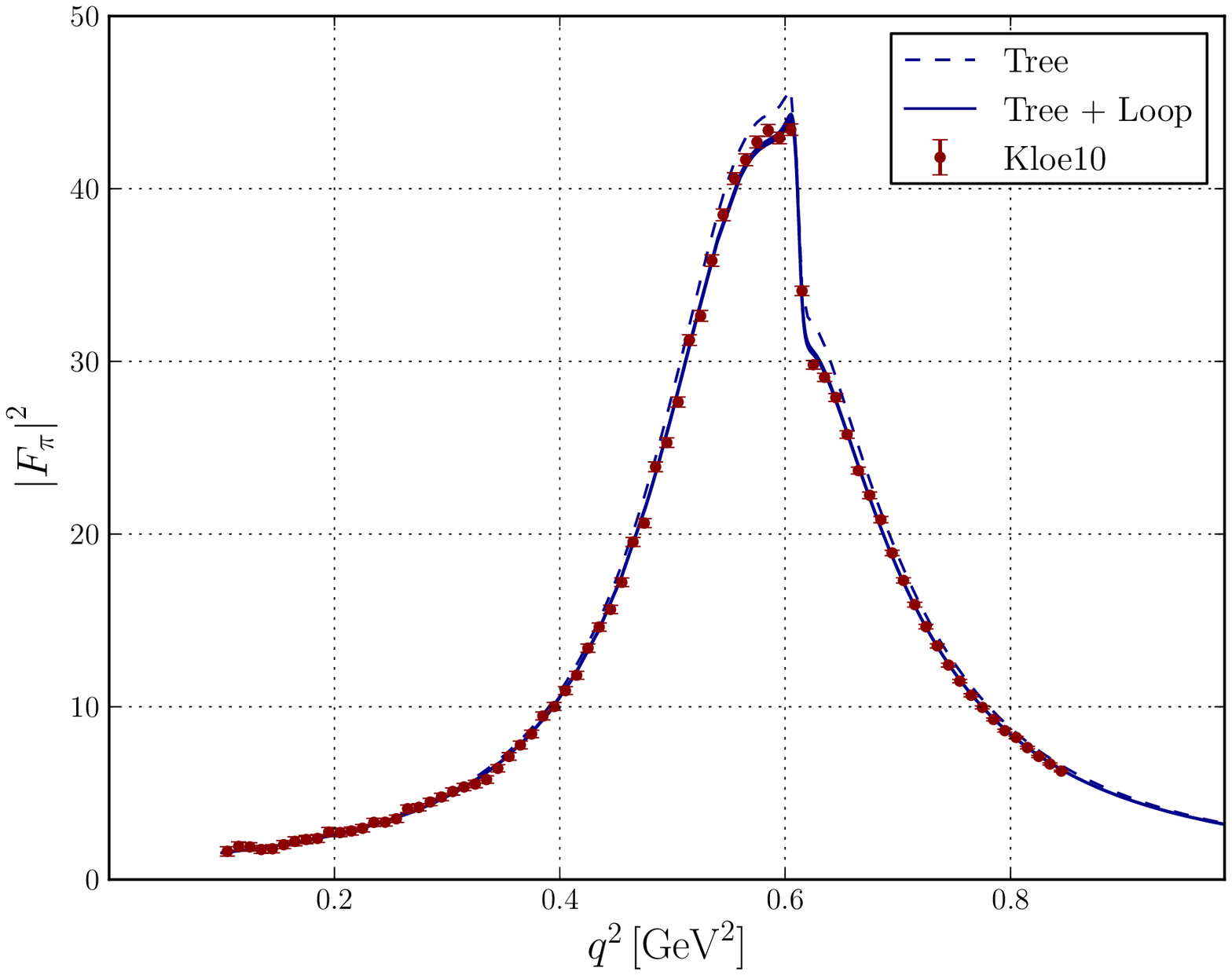,width=.4\textwidth}
  \epsfig{file=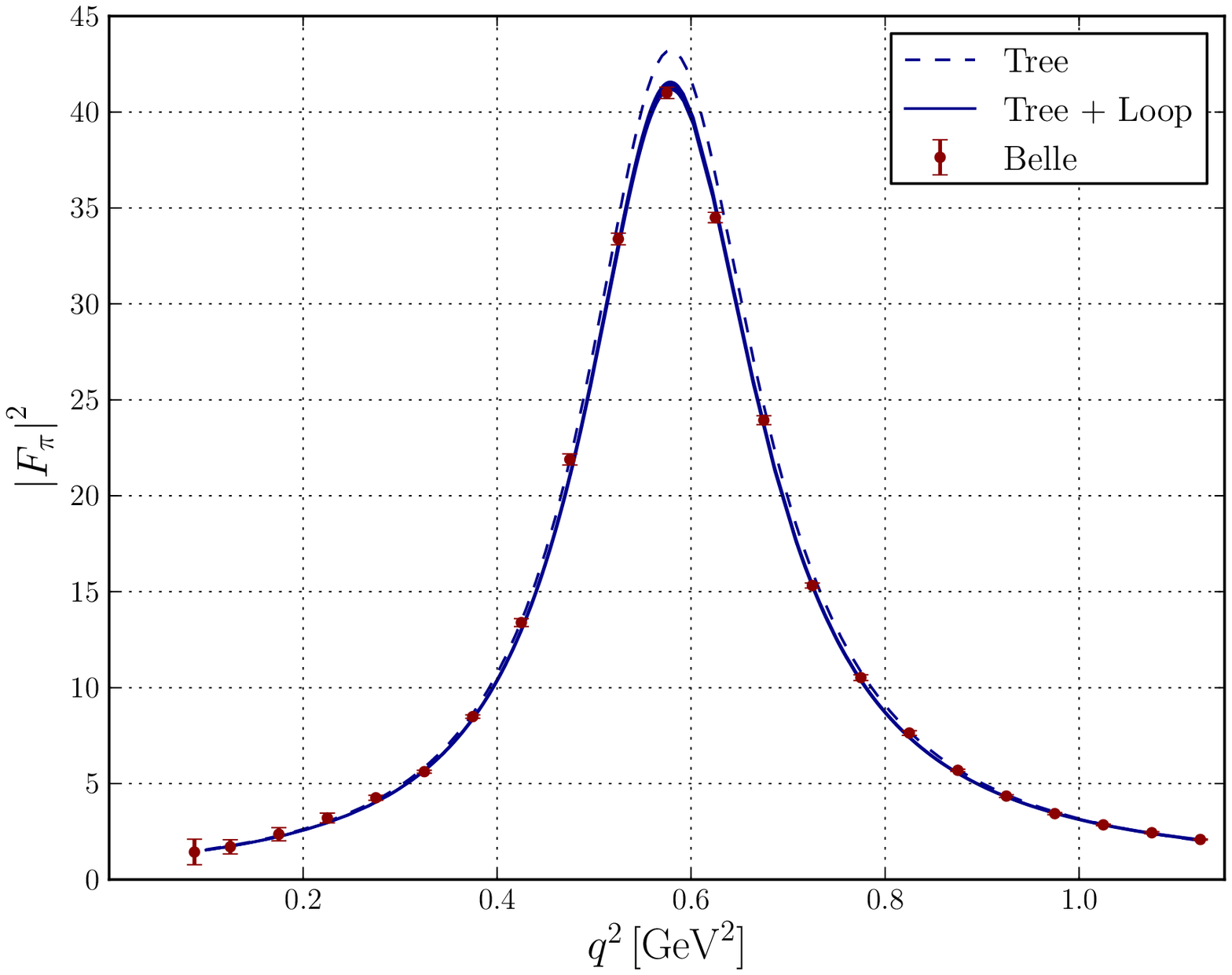,width=.4\textwidth}\epsfig{file=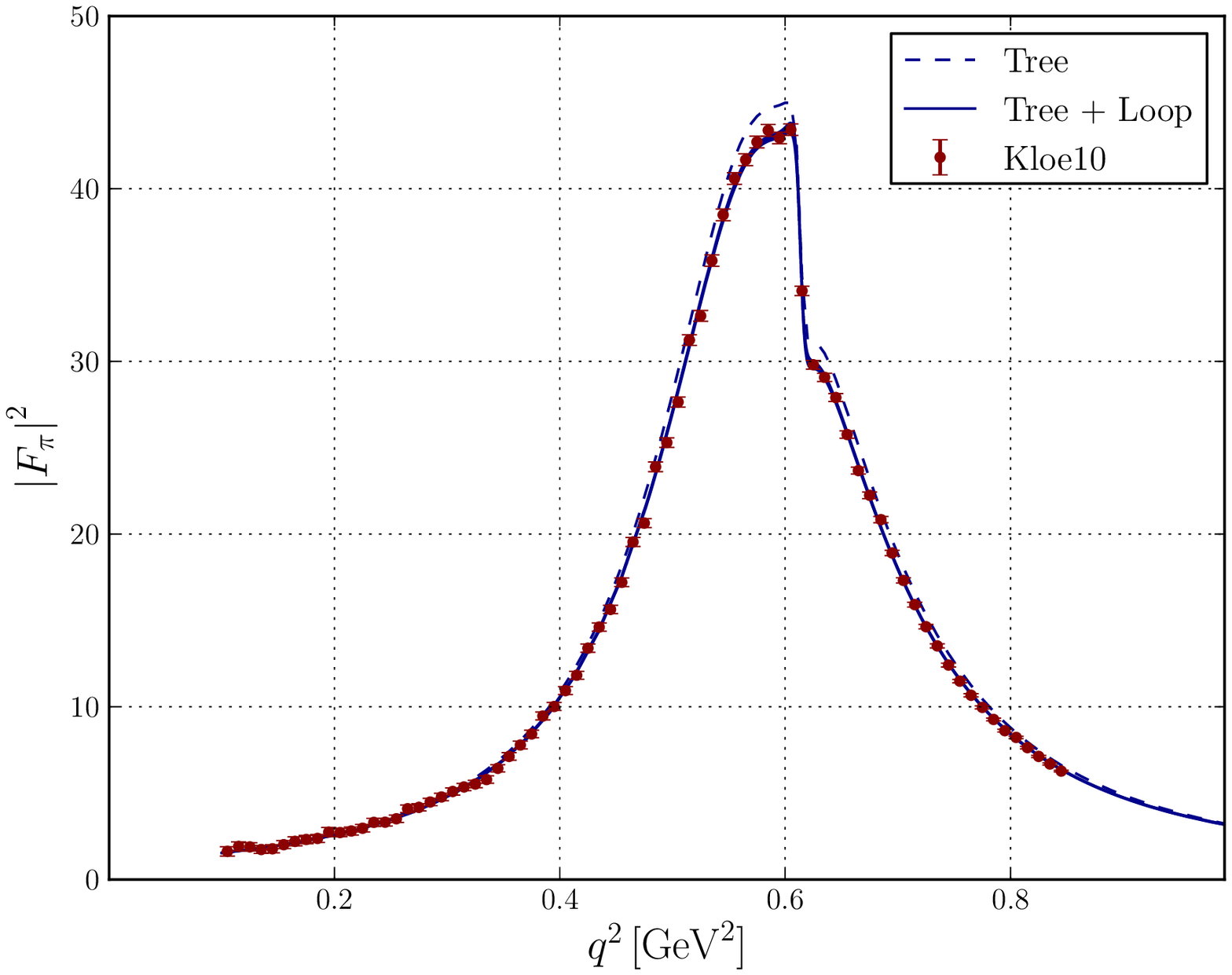,width=.4\textwidth}
  \caption{\label{fits} Fits to the pion form factor data extracted from $\tau$ decay
  \cite{Fujikawa:2008ma} (left)
  and $e^+e^-$ scattering \cite{Ambrosino:2010bv} (right). The systematic and statistical errors were
  added in quadrature for the $\tau$ decay data. In the first (second) row the
  $\omega$ mass is fixed (floating).}

  \end{figure}

\section{Conclusions}

   We have calculated the vector form factor of the pion in the framework of chiral
EFT with vector mesons included as dynamical degrees of freedom.
   To renormalize the loop diagrams, we applied the CMS.
   Within this renormalization scheme, the given EFT has a consistent power counting.
   By fitting the available parameters of the Lagrangian, a satisfactory description of the
data extracted from $\tau^-\rightarrow \nu_\tau \pi^- \pi^0$ decay has been obtained.
   On the other hand, to achieve a reasonable  accuracy in describing the form factor extracted from
$e^+ e^-\to \pi^+\pi^-$ data, it is necessary to incorporate
the $\rho^0$-$\omega$-$\gamma$ mixing.
   We included this mixing only at the tree level.
   While a satisfactory fit to the data has been obtained by fitting the mixing parameters, more work
needs to be done to incorporate the isospin-symmetry-breaking effects in a systematic fashion.
   This is subject of a future project.
   From our results we conclude that a chiral EFT with explicitly incorporated resonance states
is a promising candidate for a successful phenomenological description of data beyond the
low-energy region of ChPT.

\acknowledgments

This work was supported in part by Georgian Shota Rustaveli National
Science Foundation (grant 11/31) and
by the Deutsche Forschungsgemeinschaft (SFB/TR 16,
``Subnuclear Structure of Matter'' and SFB 1044).

\section{Appendix}

\appendix

The loop functions $A_0$, $B_0$, and $C_0$ contributing to the pion
form factor diagrams are defined as follows:
\begin{align*}
A_0(m^2)&=\frac{(2 \pi)^{4-n}}{i\,\pi^2}\,\int \frac{d^nk}{k^2-m^2+i0^+}\,,\\
B_0(p^2,m_1^2,m_2^2)&=\frac{(2 \pi)^{4-n}}{i\,\pi^2}\,
\int\frac{d^nk}{\left[k^2-m_1^2+i0^+\right]\left[(p+k)^2-m_2^2+i0^+\right]}\,,\\
\lefteqn{C_0(p_1^2,(p_2-p_1)^2,p_2^2,m_1^2,m_2^2,m_3^2)
}\\
&=\frac{(2 \pi)^{4-n}}{i\,\pi^2}\int
\frac{d^nk}{\left[k^2-m_1^2+i0^+\right]\left[(p_1+k)^2-m_2^2+i0^+\right]\left[(p_2+k)^2-m_3^2+i0^+\right]}\,,
\end{align*}
where $n$ is the space-time dimension.

\medskip

To one-loop order, the wave function renormalization constant of the pion, $1+\delta Z_\pi$, is
given by
\begin{align}
\delta Z_\pi&=\frac{1}{96F_\pi^4g^2\pi^2}
\Biggl(\frac{3(M_R^2+c_x M_\pi^2)^2}{M_R^2 M_\pi^2}
\biggl\{M_R^2[-M_\pi^2+A_0(M_\pi^2)+(M_R^2-M_\pi^2) B_0(M_\pi^2,M_R^2,M_\pi^2)]\nonumber\\
&\quad-(M_R^2-3 M_\pi^2)A_0(M_R^2)\biggr\}+[3(M_R^2+c_x M_\pi^2)-4 F_\pi^2 g^2]A_0(M_\pi^2)\Biggr).
\label{deltaZpi}
\end{align}
The contributions of the loop diagrams to the form factor read
\begin{align}
D_{l1}&=\frac{5(8F_\pi^2g^2-7 M_R^2)A_0(M_\pi^2)}{384F_\pi^4g^2\pi^2}\,,\nonumber\\
D_{l2+l3}&=\frac{3M_R^4[A_0(M_\pi^2)-A_0(M_R^2)
+(M_R^2-4 M_\pi^2)B_0(M_\pi^2,M_R^2,M_\pi^2)]}{128F_\pi^4g^2M_\pi^2 \pi^2}\,,\nonumber\\
D_{l4}&=-\frac{M_R^6}{128F_\pi^6g^4M_\pi^2\pi^2(M_R^2-q^2)(q^2-4M_\pi^2)}
\Biggl(M_R^2\Biggl\{(M_R^2-4 M_\pi^2+2 q^2)M_\pi^2 \nonumber\\
&\quad\times[2 B_0(q^2,M_\pi^2,M_\pi^2)+(2 M_R^2-4 M_\pi^2+q^2)C_0(M_\pi^2,M_\pi^2,q^2,M_\pi^2,M_R^2,M_\pi^2)]\nonumber\\
&\quad+(4 M_\pi^2-q^2)[A_0(M_R^2)-A_0(M_\pi^2)]+[16 M_\pi^4-6 q^2M_\pi^2+M_R^2(q^2-6 M_\pi^2)]\nonumber\\
&\quad\times B_0(M_\pi^2,M_R^2,M_\pi^2)\Biggr\}+(M_\pi^2 q^2-4M_\pi^4)A_0(M_R^2)\Biggr),\nonumber\\
D_{l5}&=\frac{M_R^2}{2304F_\pi^4g^2M_\pi^2\pi^2(M_R^2-q^2)(q^2-4 M_\pi^2)}
\,\Biggl[18M_R^4(4 M_\pi^2-q^2)A_0(M_\pi^2)\nonumber\\
&\quad+18M_R^4[-16M_\pi^4+(8 M_R^2+6 q^2)M_\pi^2-M_R^2 q^2] B_0(M_\pi^2,M_\pi^2,M_R^2)\nonumber\\
&\quad-18M_\pi^2M_R^4(-2 M_R^2+8 M_\pi^2-q^2)(2 M_R^2+q^2)C_0(M_\pi^2,M_\pi^2,q^2,M_R^2,M_\pi^2,M_R^2)
\nonumber\\
&\quad-2M_\pi^2(4M_\pi^2-q^2)(12M_R^4-8q^2M_R^2+(q^2)^2)\nonumber\\
&\quad+6(4 M_\pi^2-q^2)[M_\pi^2(10M_R^2+q^2)-3M_R^4]A_0(M_R^2)\nonumber\\
&\quad-3M_\pi^2[24M_R^6+28q^2M_R^4-18(q^2)^2M_R^2-(q^2)^3
+M_\pi^2(-64M_R^4+72q^2M_R^2+4(q^2)^2)]\nonumber\\
&\quad\times
B_0(q^2,M_R^2,M_R^2)\Biggr],\nonumber\\
D_{l6}&=\frac{5 M_R^4 A_0(M_\pi^2)}{384 F_\pi^4 g^2 \pi^2 (M_R^2-q^2)},\nonumber\\
D_{l7}&=\frac{M_R^4 (3 M_R^2-4 F_\pi^2 g^2)
[12M_\pi^2-2q^2+6 A_0(M_\pi^2)+3(4M_\pi^2-q^2)B_0(q^2,M_\pi^2,M_\pi^2)]}{2304 F_\pi^6
g^4 \pi^2(M_R^2-q^2)}\,,\nonumber\\
D_{l8}&=\frac{M_R^4 A_0(M_\pi^2)}{32 F_\pi^4 g^2 \pi^2 (M_R^2-q^2)}\,,\nonumber\\
D_{l9}&=-\frac{3M_R^4[5 M_R^2-6A_0(M_R^2)]}{128F_\pi^2\pi^2(M_R^2-q^2)^2}\,,\nonumber\\
D_{l10}&=\frac{M_R^8[2(q^2-6 M_\pi^2)-6A_0(M_\pi^2)
+3(q^2-4 M_\pi^2)B_0(q^2,M_\pi^2,M_\pi^2)]}{1152F_\pi^6g^4\pi^2(M_R^2-q^2)^2}\,,\nonumber\\
D_{l11}&=-\frac{1}{1152F_\pi^2\pi^2(M_R^2-q^2)^2}\Biggl[9M_R^6-48q^2M_R^4+20(q^2)^2M_R^2
-2(q^2)^3\nonumber\\
&\quad+6(3M_R^4+8q^2M_R^2+(q^2)^2)A_0(M_R^2)\nonumber\\
&\quad+3 (48 M_R^6+68 q^2M_R^4-16 (q^2)^2M_R^2-(q^2)^3)B_0(q^2,M_R^2,M_R^2)\Biggr].
\label{loopdiagrams}
\end{align}

\end{document}